\documentclass[aps,prb,reprint,longbibliography,floatfix]{revtex4-2}

\usepackage[T1]{fontenc}
\usepackage[utf8x]{inputenc}

\usepackage[american]{babel}

\usepackage{graphicx}
\usepackage[export]{adjustbox}[2011/08/13]
\usepackage{amsmath}
\usepackage{microtype}
\usepackage[normalem]{ulem}
\usepackage{siunitx}
\usepackage{tikz}

\newcommand{\mytitle}{Effect of wear particles and roughness on
  nanoscale friction}

\usepackage{hyperref} %
\hypersetup{
  final,
  pdfborder={0 0 0},
  pdftitle={\mytitle},
  pdfauthor={Brink, Milanese, Molinari},
  colorlinks=true,
  urlcolor=blue,
  linkcolor=blue,
  citecolor=blue
}

\usepackage{xcolor}
\definecolor{green}{rgb}{0., 0.45, 0.}

\usepackage{pdfpages}
\makeatletter
\AtBeginDocument{\let\LS@rot\@undefined}
\makeatother

\makeatletter
\newcommand*{\balancecolsandclearpage}{%
  \close@column@grid
  \cleardoublepage
  \twocolumngrid
}
\makeatother
\begin{document}
\frenchspacing

\title{\mytitle}

\author{Tobias Brink}
\email{t.brink@mpie.de}
\altaffiliation{Present address: Max-Planck-Institut f\"ur
  Eisenforschung GmbH, Max-Planck-Stra\ss{}e 1, 40237 D\"usseldorf,
  Germany}
\affiliation{Civil Engineering Institute and Institute of Materials
  Science and Engineering, \'Ecole polytechnique f\'ed\'erale de
  Lausanne (EPFL), Station 18, CH-1015 Lausanne, Switzerland}

\author{Enrico Milanese}
\altaffiliation{Present address: Department of Earth, Atmospheric, and
  Planetary Sciences, Massachusetts Institute of Technology,
  Cambridge, MA, USA}
\affiliation{Civil Engineering Institute and Institute of Materials
  Science and Engineering, \'Ecole polytechnique f\'ed\'erale de
  Lausanne (EPFL), Station 18, CH-1015 Lausanne, Switzerland}

\author{Jean-François Molinari}
\affiliation{Civil Engineering Institute and Institute of Materials
  Science and Engineering, \'Ecole polytechnique f\'ed\'erale de
  Lausanne (EPFL), Station 18, CH-1015 Lausanne, Switzerland}

\date{\today}

\begin{abstract}
  Frictional contacts lead to the formation of a surface layer called
  the third body, consisting of wear particles and structures
  resulting from their agglomerates. Its behavior and properties at
  the nanoscale control the macroscopic tribological performance. It
  is known that wear particles and surface topography evolve with time
  and mutually influence one another. However, the formation of the
  mature third body is largely uncharted territory and the properties
  of its early stages are unknown. Here we show how a third body
  initially consisting of particles acting as roller bearings
  transitions into a shear-band-like state by forming adhesive bridges
  between the particles. Using large-scale atomistic simulations on a
  brittle model material, we find that this transition is controlled
  by the growth and increasing disorganization of the particles with
  increasing sliding distance. Sliding resistance and wear rate are at
  first controlled by the surface roughness, but upon agglomeration
  wear stagnates and friction becomes solely dependent on the real
  contact area in accordance with the plasticity theory of contact by
  Bowden and Tabor.
\end{abstract}

\maketitle

\newcounter{supplfigctr}
\renewcommand{\thesupplfigctr}{S\arabic{supplfigctr}}
{\refstepcounter{supplfigctr}\label{fig:tribolayer-thickness}}
{\refstepcounter{supplfigctr}\label{fig:random-particle-placement}}
{\refstepcounter{supplfigctr}\label{fig:welding-metal-sims}}
{\refstepcounter{supplfigctr}\label{fig:asperity-level-wear-law}}
{\refstepcounter{supplfigctr}\label{fig:friction-wear-big-particles}}
{\refstepcounter{supplfigctr}\label{fig:wear-particle-shape}}
{\refstepcounter{supplfigctr}\label{fig:smaller-bulk}}
{\refstepcounter{supplfigctr}\label{fig:ramp-down-up}}
{\refstepcounter{supplfigctr}\label{fig:non-rigid}}
\newcommand{\supplref}[2][]{\ref*{#2}#1 in the Supplemental Material
  \cite{supplmat}}

\section{Introduction}

\begin{tikzpicture}[remember picture,overlay]
  \node [anchor=north west, font=\footnotesize, align=left,
         text width=7.05in, xshift=0.75in, yshift=0.75in, inner xsep=0pt]
        at (current page.south west)
        {Published in:\\
         \href{https://doi.org/10.1103/PhysRevMaterials.6.013606}
              {T.~Brink, E.~Milanese, and J.-F.~Molinari,
               Phys.~Rev.~Mater.~\textbf{6}, 013606 (2022)}
         \hfill
         DOI: \href{https://doi.org/10.1103/PhysRevMaterials.6.013606}
                   {10.1103/PhysRevMaterials.6.013606}
         \\
         \copyright{} 2022 American Physical Society.};
\end{tikzpicture}%
When surfaces come into frictional contact, they can undergo
structural, morphological, and/or chemical changes. A surface layer
forms, the so-called third body (or ``gouge'' in geophysics), which
consists of wear particles detached during sliding, consolidated
debris, and an altered subsurface layer \cite{godet1984third}. While
this is often a nanometer or micrometer-sized object, it impacts and
controls friction and wear at the macroscopic scale
\cite{godet1984third, brodsky2011faults, harris2015wear,
  hintikka2017third, hsia2020wear}. A significant difficulty in the
study of the third body is that it forms at an interface that is
generally inaccessible \textit{in situ} \cite{rabinowicz1995friction,
  Jacobs2017a}. Experiments are thus only amenable to post-mortem
analysis without a detailed understanding of the dynamics and
mechanisms. Even the simplest case of adhesive wear of unlubricated
surfaces, which we are concerned with in the present work, is due to a
complex interplay of different mechanisms. We nevertheless know,
partially thanks to computer simulations that allow \textit{in silico}
investigation on the atomistic length scale, that wear particles
appear and evolve in tandem with the surface roughness in inseparably
entangled processes \cite{rabinowicz1995friction, yuan2008surface,
  deng2019simple, Milanese2019, Milanese2020a}. In a typical two-body
contact, third body formation commences by the detachment of wear
particles based on the morphology of the contacts between the surfaces
\cite{godet1984third, harris2015wear, hintikka2017third, hsia2020wear,
  Aghababaei2016, Aghababaei2017, Brink2019, Aghababaei2019a}. These
particles work the surface and interact with each other, possibly
agglomerating \cite{godet1984third, harris2015wear, Milanese2019,
  Milanese2020, Milanese2020a}. Many atomistic simulations addressing
third-body rheology have been reported in the literature, but they
have either not been able to capture wear particle formation
\cite{Sorensen1996, Spijker2011, Pastewka2011, Stoyanov2013,
  Romero2014}, neglected surface roughness \cite{Mo2009, Yang2016,
  Sharp2016, Sharp2017}, used two-dimensional models
\cite{Aghababaei2019a, Milanese2019, Milanese2020a}, restricted the
movement of wear particles \cite{eder2015applicability}, or used only
a single wear particle \cite{Aghababaei2019a, Milanese2019,
  Milanese2020a}. Here, we present molecular dynamics (MD) computer
simulations with a multi-particle, three-dimensional setup with fully
plastically-deformable, rough surfaces.

\section{Methods}

All simulations in the present work were molecular dynamics
simulations performed using the software \mbox{\textsc{lammps}}
\cite{Plimpton1995} with GPU-accelerated potentials \cite{Brown2011,
  Brown2013}.

\subsection{Model material}

The formation of wear particles has a size dependence that is a
function of material properties, where asperity--asperity contacts
below a critical size $d^*$ cannot lead to the detachment of particles
\cite{Aghababaei2016}. The relevant length scales for wear are smaller
for hard and brittle materials. Therefore we chose to use a
silicon-like model material, which has been modified to be more
brittle than real silicon \cite{Stillinger1985, Holland1998erratum,
  Brink2019} in order to be able to observe wear particle interactions
under sustained rolling conditions. The material was nanocrystalline
with average grain sizes of \SI{3}{nm} to make it more isotropic and
to accelerate the wear process (see Appendix~\ref{sec:mater-choice}
for tests and comparisons with other material choices).

The material used for the main part of our work has a diamond crystal
structure with a lattice constant of \SI{0.543}{nm}. Using a pure
shear simulation of the bulk material with an engineering shear rate
of \SI{e8}{s^{-1}}, we found a shear strength of $\tau =\SI{6.6}{GPa}$
and estimated the hardness as
$H \approx 3\sqrt{3}\tau = \SI{34.5}{GPa}$. The material has a Young's
modulus of \SI{149.5}{GPa}, a Poisson ratio of $0.23$, and a shear
modulus of $G = \SI{60.6}{GPa}$. The single crystal has a $\{111\}$
surface energy of $\gamma = \SI{1.6}{J/m^2}$, which we take to be
approximately half the fracture energy due to the brittleness of the
material. Wear particles are expected to form at asperities with sizes
exceeding a diameter of \cite{Aghababaei2016}
$d^* \approx 12 \gamma G \tau^{-2} = \SI{27}{nm}$. Asperities and
contacts with a diameter below $d^*$ will not detach during sliding,
but only deform plastically \cite{Aghababaei2016}. The value of $d^*$
for a fully amorphous system would be roughly half of that
\cite{Brink2019} and we thus expect the value for the nanocrystal to
lie in between those two values, since the grain boundaries are
expected to have lower fracture resistance.

\subsection{Simulation setup}

To prepare the sliding simulations, two first bodies consisting of the
bulk material were prepared with periodic boundary conditions in $x$
and $y$ direction. In $z$ direction, we used pre-worn surfaces to
speed up the running-in process (see
Appendix~\ref{sec:setup-choices}).  The initial third body consists of
several, independent rigid particles with the same interatomic
potential as the bulk material. We chose not to model the particle
formation process for considerations of computational efficiency,
since it has been investigated before in detail \cite{Aghababaei2016,
  Aghababaei2017, Brink2019, Aghababaei2019a, Milanese2019,
  Milanese2020, Milanese2020a}. For the main simulations, we used 16
particles in the shape of rhombicuboctahedra with a diameter of about
\SI{8}{nm} arranged on a regular grid, which is roughly on the order
of the minimal particle diameter $d^*$ at its formation (see
Appendix~\ref{sec:setup-choices} for different choices of initial
particle shape and placement).

The two first bodies each had a size of \SI{58x58x11}{nm} (around
3,200,000 atoms for the whole simulation), where the topmost and
lowermost layers of width \SI{0.4}{nm} in $z$ direction were treated
as rigid boundaries to apply normal force and sliding
displacement. Possible size effects are discussed in
Appendix~\ref{sec:setup-choices}. The energy of the system was
minimized to avoid large forces due to any closely-spaced atoms that
might have appeared during the preparation. The systems were then
equilibrated with a Nosé--Hoover thermostat at \SI{300}{K} and
barostats at ambient pressure in $x$ and $y$ direction. The target
normal force $F_N$ was applied on the boundaries in $z$
direction. This equilibration was performed for \SI{1}{ns} with a time
integration step of \SI{1}{fs}.

After equilibration, sliding simulations with a velocity of
\SI{20}{m/s} were performed over a given sliding distance $s$. The
sliding direction was chosen to be \ang{8.5} off the $x$ direction to
avoid wear particles moving along the same groove during the whole
simulation due to the periodic boundary conditions. The sliding
velocity was initially applied to the whole upper body and
continuously enforced at the top boundary throughout the
simulation. The lower boundary was kept fixed. The reaction forces in
sliding direction at the boundaries then equal the friction force
$F_T$. To keep the simulations isothermal at room temperature,
Langevin thermostats with a damping constant of \SI{0.01}{ps} were
applied in the regions next to the top and bottom boundaries over a
width of \SI{0.4}{nm}. The center-of-mass velocity of the top
thermostat region was subtracted before applying the thermostat in
order to avoid an artificial drag force.

We also performed some verification simulations to investigate the
influence of sliding velocity and the choice of rigid particles
instead of plastically deformable ones. This is described in
Appendix~\ref{sec:verif-sims}.

\subsection{Surface reconstruction and characterization}

In order to be able to cleanly delineate the surface from the third
body, the wear particles must be separated from the first bodies. For
this, we simply fixed the rigid wear particles in space and applied a
velocity of \SI{20}{m/s} in $z$ direction to each boundary to pull the
bodies apart. In some cases with larger wear volume, wear fragments
not belonging to the rigid particles stuck to one of the surfaces, in
which case we also fixed a region of width \SI{1}{nm} in the center of
the tribolayer. At the end of the separation process, the atomic
positions were minimized with regard to the potential energy to remove
the frozen thermal vibrations for the analysis.

Since atoms in MD simulations are modeled as mass points, the notion
of a surface is ill defined without further specifications. Here, we
used a surface mesh generation algorithm \cite{Stukowski2014} as
implemented in \textsc{ovito} \cite{Stukowski2010} to find atoms that
belong to a surface of the volume that is impenetrable to a virtual
probe sphere. We chose the radius of the probe sphere to roughly
correspond to the average
second-neighbor distance of \SI{3.85}{\angstrom} for silicon.

As a simple parameter for describing the surface roughness, we used
the rms of heights
$h_\text{rms} = \langle (z_i - \langle z_i\rangle)^2\rangle^{1/2}$,
where $z_i$ is the $z$ component of the position of the surface atoms
and $\langle\ldots\rangle$ is the arithmetic mean. Both first bodies
were made of the same material, so they quickly attained the same
value of $h_\text{rms}$, even during the running-in phase. We
therefore use the arithmetic mean of the values for both surfaces from
here on.

\subsection{Estimation of wear volume and third body thickness}

The wear volume $V$ was estimated from the separated third body in the
middle of the simulation box. For this, we calculated the average
atomic volume in the tribolayer by Voronoi tessellation
\cite{Brostow1998} as $\Omega = \SI{21.292e-3}{nm^3}$. The wear volume
is thus defined as the number of atoms in the third body (excluding
the initially inserted wear particles) times $\Omega$.

We defined the distance $g$ between the surfaces of the two first
bodies---which we also define as the thickness of the third body and
which is approximately equal to the diameter of the wear
particles---with the help of a density profile along the $z$ direction
of the non-separated simulation boxes. Since the density in the
tribolayer is lower, we defined the average surfaces of the first
bodies by taking the half height between minimum and maximum density
on both sides of the third body [inset of
Fig.~\supplref[(a)]{fig:tribolayer-thickness}]. We take the distance
between these two points as the thickness $g$ of the third body. It is
also possible to verify the wear volume by integrating over the
density data in the tribolayer. This method mostly agrees with the
first method described above
[Fig.~\supplref[(b)]{fig:tribolayer-thickness}].

\subsection{Contact area}

The real contact area $A_c$ between the third body and the first bodies was
computed by marking the atoms belonging to the third body (obtained
via the separation method outlined above) and considering all atoms in
the first bodies that are closer than \SI{0.3}{nm} to the third body
to be in contact. The cutoff was chosen as the first minimum in the
pair distribution function. To convert the number of contacting atoms
to an area, we used a typical surface, calculated its area and the
number of surface atoms using surface mesh generation
\cite{Stukowski2014}, and arrived at a surface area per surface atom
of \SI{0.1053}{nm^2}.

\section{Results}
\subsection{The third body}

\begin{figure}[b]
  \centering
  \includegraphics[center]
    {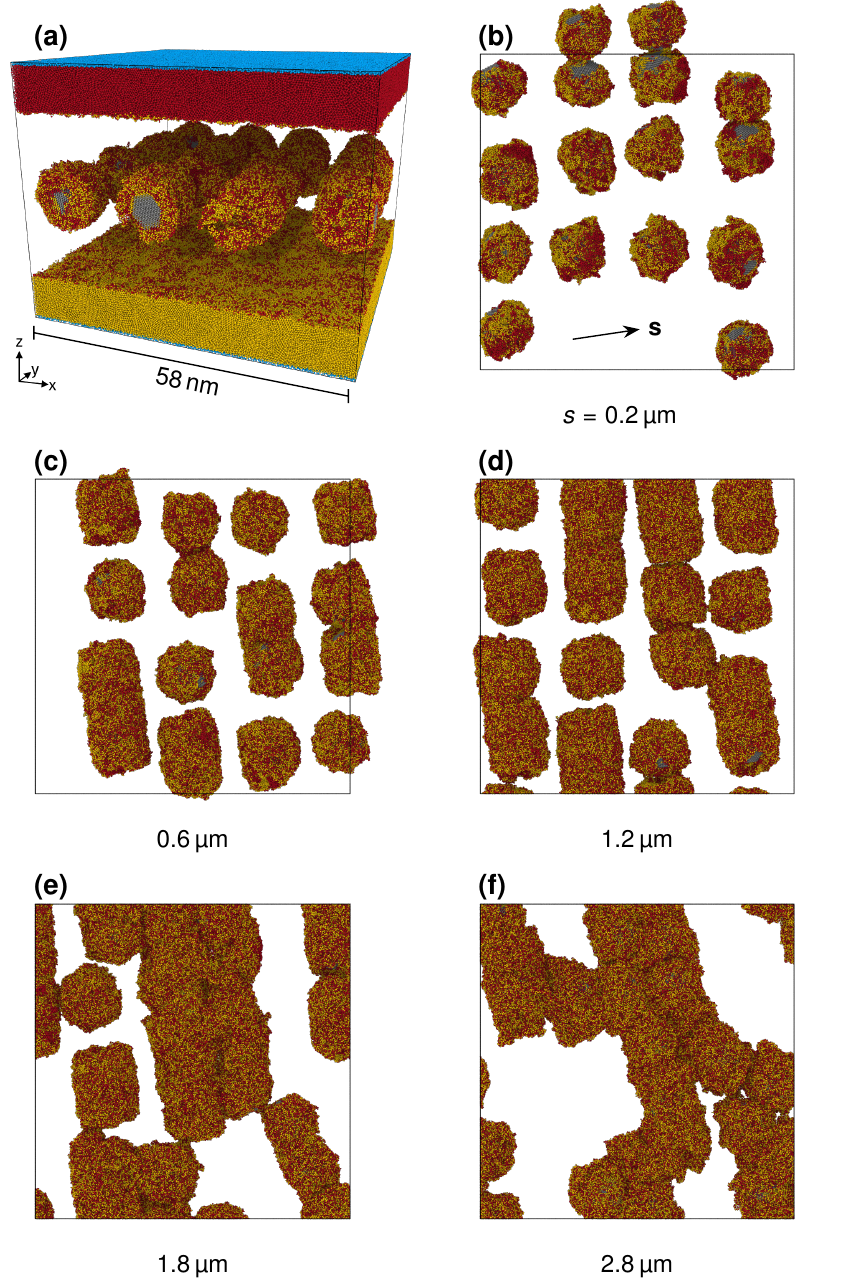}
    \caption{Early stages of the third body. (a) Particles were
      introduced between two surfaces and subjected to sliding
      motion. Panels (b)--(f) show only the third body as it evolves
      from spherical particles (b), to cylindrical rollers (c)--(d),
      to a porous, shear-band-like state (e)--(f). Atoms are colored
      according to the body they originally belonged to, namely top
      surface (red), bottom surface (yellow), third body (gray), or
      boundary (blue). The box indicates the periodic boundaries and
      atoms are shown outside if it makes the shape of the particle
      clearer. The sliding direction is indicated in panel (b).}
  \label{fig:snapshots-3rd-body-evolution}
\end{figure}

The introduction of rigid wear particles between two bodies in
relative sliding motion
[Fig.~\ref{fig:snapshots-3rd-body-evolution}(a)] first leads to a
relatively short running-in phase, in which the surface morphology
evolves quickly and the particles become coated with material picked
up from both surfaces [Fig.~\ref{fig:snapshots-3rd-body-evolution}(b)
and video \texttt{run-in.webm} in the companion dataset
\cite{Brink2021zenodo}]. Due to this coating, the wear process
resembles the adhesive wear case with comparable material properties
in all three bodies \cite{rabinowicz1995friction}, despite the
artificial rigidity of the initial particles (see
Appendix~\ref{sec:verif-sims} for a detailed discussion of the
difference between rigid and plastically deformable particles). The
surfaces of the two first bodies become amorphous up to a depth of
around 1 to \SI{3}{nm}. In the common scenario of adhesive wear of
initially bare surfaces, this phase would instead be dominated by the
formation of wear particles \cite{Aghababaei2016, Aghababaei2017,
  Brink2019, Milanese2019, Milanese2020}, which we bypassed here.

As expected from two-dimensional modeling and theory
\cite{Milanese2019, Milanese2020a, Milanese2020}, the spherical
particles continue to grow. In three dimensions, though, the easiest
growth direction is into the empty space lateral to the sliding
direction, leading to the formation of cylindrical rollers illustrated
in Fig.~\ref{fig:snapshots-3rd-body-evolution}(b)--(d) and the video
\texttt{rolling.webm} in the companion dataset
\cite{Brink2021zenodo}. Such cylindrical debris particles have been
observed experimentally in brittle and quasi-brittle materials
\cite{Mizuhara1992, Zanoria1993, zanoria1995effects,
  zanoria1995formation, Xu2000, reches2010fault, Hayashi2010,
  nakamura2012amorphization, Chen2017, chen2017powder}, and their
appearance relates to a weakening of the interface (i.e.\ a drop in
the friction coefficient) \cite{zanoria1995formation,
  chen2017powder}. In rock experiments \cite{Chen2017}, the interface
strengthens upon further sliding and then weakens again. This
evolution is attributed to the formation and full development of a
layer where shear localizes. Yet the transition from powder rolling to
shear localization is still unclear.

\begin{figure}[b]
  \centering
  \includegraphics[center]{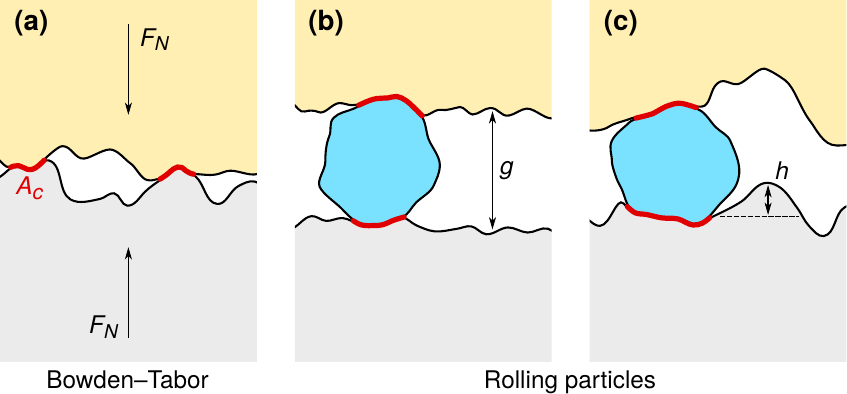}
  \caption{Influence of roughness on friction. (a) In Bowden and
    Tabor's model, friction is due to the real contact area
    $A_c \propto F_N$. The proportionality is a consequence of the
    fractal nature of the surface roughness, where more asperities
    come into contact with increasing load.  (b) When the system
    includes wear particles, this simple proportionality is
    insufficient to explain the friction force, which now also depends
    on the rolling friction of the particles. Nevertheless, there is
    still a contact area between the third-body particle and the two
    first bodies, which can lead to friction due to adhesive
    forces. (c) At least on the nanoscale, the roughness may be large
    compared to the particle size and therefore increase friction by
    presenting obstacles in the form of surface asperities to the
    smooth rolling motion of the particles. Configurations with large
    asperities should be unstable, since the obstacles are worn off
    (or the particle is reincorporated into the surface). The
    influence of roughness---i.e., the asperity heights $h$---has to
    be related to the particle size, here approximated by the
    thickness $g$ of the third body.}
  \label{fig:roughness-sketch}
\end{figure}

\begin{figure*}
  \centering
  \includegraphics[]{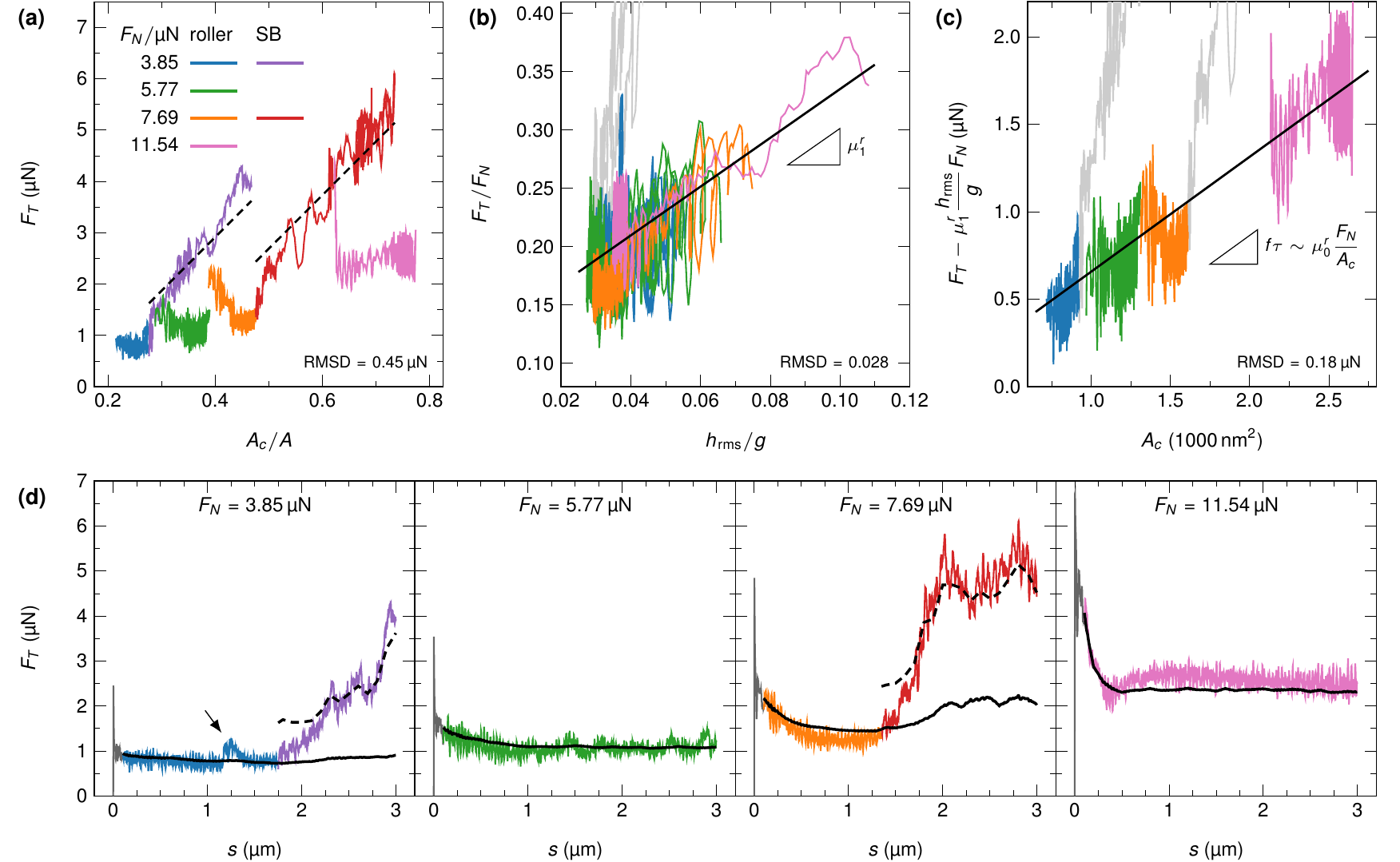}%
  \caption{Friction force as a function of contact area, roughness,
    and sliding distance. (a) The friction force plotted over the
    relative contact area reveals that a direct linear dependence
    between the two quantities only exists in the shear-band-like
    regime (SB), while the friction force of the cylindrical rollers
    does not vary with the contact area for a given normal load, or
    even seems to be slightly negatively correlated. The dashed lines
    represent fits to the friction law in
    Eq.~\ref{eq:sb-friction}. (b) The friction of the rollers is
    instead correlated with the surface roughness divided by the
    third-body thickness, in accordance with
    Eq.~\ref{eq:rolling-friction} (solid black line). The light gray
    lines represent the data from the shear-band-like regime, for
    which the roughness data is much less reliable. (c) Subtracting
    the roughness-dependent term of Eq.~\ref{eq:rolling-friction} from
    the frictional force reveals a linear relation between the
    remaining partial friction force and the real contact area (solid
    black line). This explains the apparent, partially negative
    correlation between $F_T$ and $A_c$: When the contact area
    increases after running-in, it is counteracted by the roughness
    decreasing at the same time. The slope is indirectly related to
    the friction coefficient $\mu_0^r$ of a flat surface. (a)--(c) The
    root-mean-square deviation (RMSD) is provided as a goodness-of-fit
    measure. (d) The friction laws in this work
    (Eqs.~\ref{eq:rolling-friction}, solid lines and
    Eq.~\ref{eq:sb-friction}, dashed lines) fit well to the friction
    force data plotted over sliding distance. The arrow shows a point
    in time when two cylinders temporarily stick together, leading to
    the transient friction peak when the pure rolling motion is
    disturbed.}%
  \label{fig:friction}%
\end{figure*}

In our simulations of the brittle model material, the persistence of
the regime of cylindrical rollers is contingent upon either a regular
spacing of the cylindrical particles or large distances between
them. Figure~\ref{fig:snapshots-3rd-body-evolution}(e)--(f) shows that
interactions between the particles lead to agglomeration and a
breakdown of the rolling regime. A very porous, shear-band-like state
emerges. If our simulations start with wear particles arranged on a
regular grid, this shear-band-like state is delayed and sometimes does
not occur within the time scale of the simulations. If the particles
are initially placed randomly instead, interactions between them
destabilize the rolling regime immediately
(Fig.~\supplref{fig:random-particle-placement} and video
\texttt{sb-like-regime.webm} in the companion dataset
\cite{Brink2021zenodo}).

We tried to reproduce this evolution also with realistic, metallic
materials, such as Cu, Al, Ni, a high-entropy alloy, and a metallic
glass, but the friction coefficient quickly reaches or exceeds a value
of 1.0 (see Appendix~\ref{sec:mater-choice} and
Fig.~\supplref{fig:welding-metal-sims}). These systems react by
scratching due to the abrasive particles and quickly weld. This is on
one hand due to their lower hardness and the lack of bond
directionality, which lead to much higher adhesion and plasticity,
thereby suppressing the rolling of wear particles more easily. On the
other hand, the typical size of wear particles due to adhesive wear is
expected to be much larger in metals than can be simulated using MD
\cite{Aghababaei2016}. Nevertheless, we are not aware of any work that
observes cylindrical particles acting as roller bearings in
metals. The following sections thus focus on the brittle model
material simulations.

\begin{figure*}
  \centering
  \includegraphics[center]{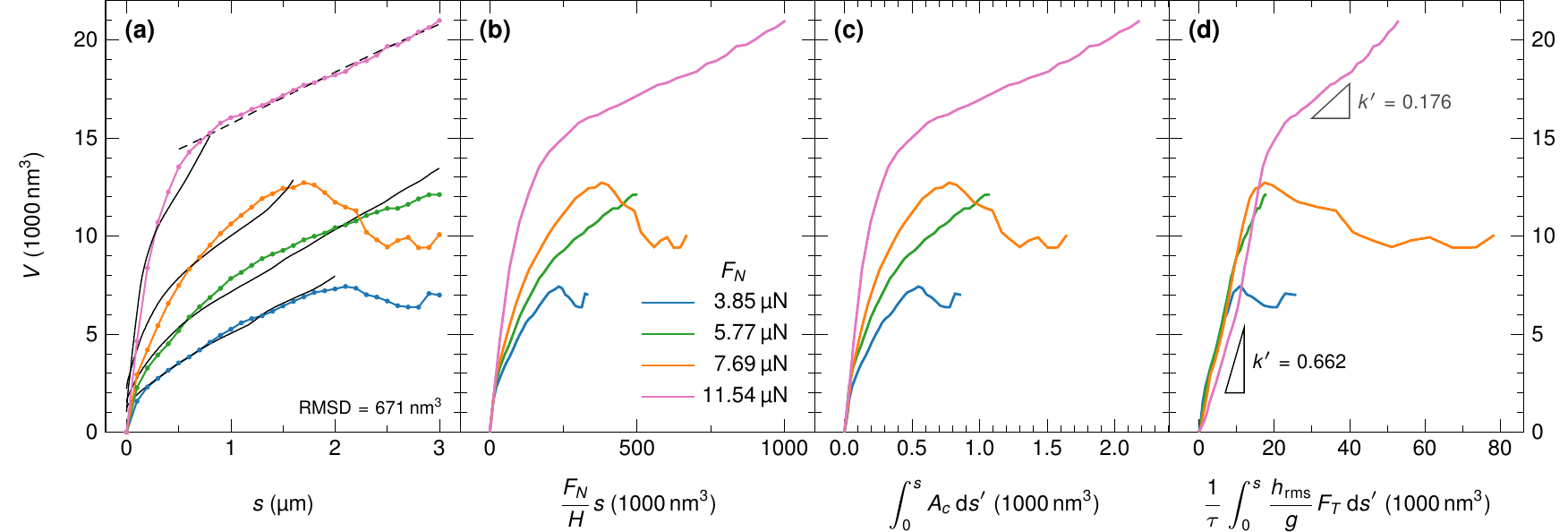}
  \caption{(a) Cumulative wear volume as a function of sliding
    distance for different normal loads. Parts of the wear volume data
    are linearly dependent on the sliding distance in the steady-state
    regime. The slopes, representing the wear rates, though, seem to
    be comparable for different normal loads. (b) By plotting the wear
    volume as a function of $F_Ns/H$ as in Archard's wear equation,
    the slopes should represent the wear coefficient $k$, which would
    be independent of $F_N$ in Archard's model. The data shows that
    this is not the case: the slopes are different and the data do not
    collapse onto a single curve. (c) In the original derivation of
    this equation, it is $F_N/H = A_c$. Therefore we also plot the
    wear volume as a function of the integral over a direct
    measurement of the contact area, but do not obtain a better
    correlation. (d) However, if we replace the normal load $F_N$ by
    the tangential friction force $F_T$ modified by a roughness term
    $h_\text{rms} / g$, we obtain the same linear relation for all
    curves including the running-in phase---although excluding the
    low-wear, shear-band-like regime---with a constant $k' = 0.662$
    (see Eq.~\ref{eq:wear}). Note that the cylinders in the simulation
    with $F_N = \SI{11.54}{\micro\newton}$ start spanning across the
    periodic boundaries at $s = \SI{0.5}{\micro\meter}$, forcing a
    rolling direction at an angle to the sliding direction. This adds
    a shear component which leads to bidirectional exchange of
    material between first and third bodies and reduces the efficiency
    of the wear process to $k' = 0.176$. The resulting best fits are
    also shown in (a) as black lines (solid line for $k' = 0.662$ and
    dashed line for $k'=0.176$) together with the RMSD as a
    goodness-of-fit measure for the solid lines.}
  \label{fig:wear}
\end{figure*}

\subsection{Friction}

In the framework of Bowden and Tabor \cite{Bowden1939}, the frictional
force $F_T$ arises as a result of the shear strength $\tau$ of
adhesive contacts between two bodies as $F_T = \tau A_c$. The real
contact area $A_c$ is smaller than the macroscopic, apparent contact
area $A$ and is proportional to the normal load $F_N$
[Fig.~\ref{fig:roughness-sketch}(a)]. At small scales, part of the
real contact area is also due to adhesion. This results in the typical
Amontons--Coulomb friction law \cite{coulomb1821theorie}, where
$F_T = \mu F_N + F_\text{adh}$ with $\mu$ being the friction
coefficient, a proportionality constant, and $F_\text{adh}$
representing the adhesive contribution at zero normal load. Previous
works found that this model is valid even at the nanoscale
\cite{Mo2009}, but third bodies or significant surface roughness were
not considered.  Simulations of abrasive particles with fixed
positions in space grinding a moving metal surface (resembling for
example polishing with sandpaper) also recovered the Bowden--Tabor
relation of friction to contact area at the nanoscale
\cite{eder2015applicability}, but did not consider the evolution and
movement of the particles. In contrast, later macroscopic model
experiments with mobile, elastic third-body particles and surfaces, as
well as low adhesion, highlight the importance of the ratio of surface
roughness to wear particle size rather than contact area
\cite{deng2019simple}.

In our case, we can observe the effects of plastic deformation, wear,
and strong adhesion that are expected to dominate the response at the
nanoscale. Figure~\ref{fig:friction}(a) shows that the friction force
has no clear dependence on the contact area in the rolling
regime. Indeed, the same friction force is observed for whole ranges
of contact area for a given $F_N$. The linearity between $A_c$ and
$F_T$ is recovered when entering the shear-band-like regime.

If the friction force for a given normal load is constant or even
slightly negatively correlated with the contact area, another
parameter must be modifying the response. It stands to reason
that---given a constant real contact area---friction on a rougher
surface should be higher. We expect this to be in relation to the size
of the rolling particles (approximated by the third body thickness
$g$). Here, we introduce a relative surface roughness, expressed in
terms of the rms of heights $h_\text{rms}$ divided by $g$, with the
reasoning being that asperities with heights of the same order as the
particle diameter represent obstacles, while much smaller asperities
barely influence the rolling motion
[Fig.~\ref{fig:roughness-sketch}(b)--(c)]. Figure~\ref{fig:friction}(b)
shows that this variable is a good predictor of the friction force in
the rolling regime, leading to a friction law of the form
\begin{equation}
  \label{eq:rolling-friction}
  \dfrac{F_T}{F_N}
  = \mu_0^\mathrm{r} + \mu_1^\mathrm{r} \dfrac{h_\text{rms}}{g}
\end{equation}
with fit parameters $\mu_0^\mathrm{r} = 0.13$ and
$\mu_1^\mathrm{r} = 2.09$. The first term resembles Amontons--Coulomb
friction. In Fig.~\ref{fig:friction}(c) the frictional force minus the
roughness-dependent term of Eq.~\ref{eq:rolling-friction} is plotted
over the real contact area. There is now a proportionality between
these quantities:
$F_T - \mu_1^r F_N h_\text{rms}/g \approx f \tau A_c$ with $f =
0.1$. We can thus also write Eq.~\ref{eq:rolling-friction} as
\begin{equation}
  \label{eq:rolling-friction-B-T}
  F_T = f \tau A_c + \mu_1^\mathrm{r} \dfrac{h_\text{rms}}{g} F_N,
\end{equation}
obtaining a version of Bowden and Tabor's friction model for the first
term, although with a significantly reduced shear strength. The
proportionality between $F_N$ and $A_c$ changes during wear of the
surfaces, meaning that there is no simple relation between $f\tau$ and
$\mu_0^r$. Our results show that the effective shear strength is
expectedly lower for the rolling regime ($f\tau$ with $f < 1$) than
for the sliding contact ($\tau$) envisioned in the original model.

When transitioning to the shear-band-like third body, there is a direct
dependence on contact area [Fig.~\ref{fig:friction}(a) and
Fig.~\supplref{fig:random-particle-placement}], but not all the area
seems to participate:
\begin{equation}
  \label{eq:sb-friction}
  F_T =\tau^\text{SB} (A_c - A_0)
\end{equation}
with $\tau^\text{SB} = \SI{3.1}{GPa}$. The bulk shear strength of the
amorphous material (as expected in the third body) is around
\SI{4.9}{GPa} \cite{Brink2019}, meaning that the third body has a
lower shear strength than the bulk material, but a higher strength
than the rolling interface ($f\tau \approx \SI{0.7}{GPa}$ in
Eq.~\ref{eq:rolling-friction-B-T}). The term depending on the relative
roughness disappears together with the rolling motion of the wear
particles. The reduction of the participating contact area by $A_0$
compared to Bowden and Tabor's model is on one hand due to parts of
the third body still rolling and on the other hand due to some
uncertainty in assigning a correct contact area per atom.

Taking both regimes into account, the friction force is well
reproduced by Eqs.~\ref{eq:rolling-friction} and \ref{eq:sb-friction}
over the whole sliding distance [Fig.~\ref{fig:friction}(d)].

\subsection{Wear}

The wear behavior of the system is plotted in
Fig.~\ref{fig:wear}(a). Since no volume can be lost in the simulation
setup due to periodic boundaries, we define the wear volume as the
volume of the third body. After the running-in phase with a high wear
rate, the rolling regime exhibits a roughly constant wear rate as it
is typically also observed in macroscopic wear \cite{Reye1860,
  Rabinowicz1951,queener1965transient}. In the macroscopic case, the
wear volume $V$ is predicted by Archard's semi-empirical wear equation
\cite{archard1953contact,archard1956wear}
\begin{equation}
    \label{eq:archard}
    \frac{V_\text{Archard}}{s} = k \frac{F_N}{H},
\end{equation}
with sliding distance $s$, hardness $H$, and wear coefficient $k$. In
the present work, the wear rate is comparable between the simulations
with different normal load [slope of the curves in
Fig.~\ref{fig:wear}(a)], which means that this system does not follow
Archard's prediction. The data does not collapse onto a single curve
when plotted as a function of $F_Ns/H$, indicating a change of $k$
with normal load [Fig.~\ref{fig:wear}(b)]. In the original derivation,
the equation can also be expressed as $V/s = k A_c$, with $F_N/H$
being an approximation of the contact area. While this is able to
account for changing contact areas during sliding, it does not differ
qualitatively from the results using Eq.~\ref{eq:archard}
[Fig.~\ref{fig:wear}(c)]. A different approach is required.

In the case of wear particle formation with significant plastic
activity, it has been found that the wear volume in the
single-asperity case is related to the tangential work by
\cite{Aghababaei2017}
\begin{equation}
  \label{eq:aghababaei2017}
  V_\text{asp} = \dfrac{\int_0^sF_T \mathrm{d}s'}{\omega \tau}
\end{equation}
with $\omega$ being a shape factor close to unity (see
Fig.~\supplref{fig:asperity-level-wear-law}). Here, all dissipated
energy is used to detach the particle.

Rolling wear particles should constantly pick up volume and wear the
surface by a fracture process \cite{Milanese2020}. In our case, the
wear particles interact with asperities of the rough surface and we
consequently found that a similar approach to the asperity--asperity
interaction also applies to wear in the rolling regime:
\begin{equation}
    \label{eq:wear}
    V = \frac{k'}{\tau} \int_0^s \frac{h_\text{rms}}{g}F_T \mathrm{d}s'
\end{equation}
with $k' = 0.662$ independent of the applied normal load
[Fig.~\ref{fig:wear}(d)]. We included a measure of the relative
roughness to achieve this, which plays a similar role to the classical
wear coefficient: it describes the efficiency of the wear
process. While in the asperity--asperity collision described by
Eq.~\ref{eq:aghababaei2017} almost all of the tangential work is
expended to detach the wear particle, only part of the friction is due
to wear particle formation and growth in more complex and realistic
scenarios. We reason that the nanoscale wear events should be more
likely if the asperities present larger obstacles to the rolling
particles, and thus that $h_\text{rms}/g$ is a reasonable measure of
the wear efficiency. This assumption allows us to also capture the
running-in phase, which is characterized by a higher roughness. The
remaining empirical parameter $k'$ is close, but not equal, to unity,
indicating that the physics of the present system can be described
adequately with the proposed physical quantities in
Eq.~\ref{eq:wear}. The relative roughness $h_\text{rms}/g$ lies in the
range of $0.02$--$0.08$ in the steady rolling regime, and therefore
makes the wear much less efficient than in the case of wear particle
formation.

In the shear-band-like regime, the wear volume stagnates. While there
is a bias to pick up volume in the rolling regime
\cite{Milanese2019,Milanese2020}, matter exchanges in both directions
between surface and the shear-band-like third body. Wear damage could
still occur by growth of the shear band, although the width of the
tribolayer stagnates at least on the timescale of our simulations
(Fig.~\supplref{fig:tribolayer-thickness}). This is thus the least
efficient wear mode, since wear rates are low or zero and friction
increases drastically.

\section{Discussion}

The appearance of cylindrical rolling particles is consistent with
experimental observation in silicon/silica \cite{Mizuhara1992,
  Zanoria1993, zanoria1995effects, zanoria1995formation}, silicon
nitride \cite{Xu2000}, and rock \cite{Hayashi2010,
  nakamura2012amorphization, Chen2017, chen2017powder}
tribosystems. In these experiments, the cylindrical particles are
often found to be amorphous \cite{Mizuhara1992, zanoria1995effects,
  zanoria1995formation, Xu2000}, which is consistent with the
amorphous layers found on top of both first and third bodies in our
simulation (note that the cores of our wear particles were rigid and
therefore impossible to amorphize). In chert (a quartz rock), smooth
parts of the surface were found to exhibit cylindrical particles,
while rougher parts were free of them \cite{Hayashi2010}. We found
that the rollers wear the surface and thus have a polishing effect,
which could explain this observation. In the literature, the presence
of cylindrical rollers is consistently reported to reduce the friction
compared to cases without the rollers \cite{Zanoria1993,
  zanoria1995effects, zanoria1995formation, Chen2017, chen2017powder}.
Friction coefficients of around $\mu = 0.2$ were observed in
silicon/silica systems when rolling particles appeared
\cite{Zanoria1993, zanoria1995effects, zanoria1995formation}, which is
similar to our results after running-in. Further comparison with our
wear and friction models is not possible, as these works do not report
quantitative measurements of the surface roughness.

What can the present results tell us about the life-cycle of the third
body and how can it be applied to larger systems? Comparing
Eqs.~\ref{eq:aghababaei2017} and \ref{eq:wear}, it is clear that most
of the wear volume should be due to wear particle formation. After
forming wear particles, though, a relatively low-wear, low-friction
regime of roller bearings is entered. The stability of this regime
depends on both the proximity of the cylindrical particles, as
interaction between them disturbs the rolling motion, and on their
size compared to the roughness. If asperities are bigger than the
particles, which is plausible in more general situations due to the
self-affine scaling of roughness with system size
\cite{power1988roughness, persson2004nature, renard2013constant}, new
wear particles are expected to form concurrently and therefore also
destabilize the rolling regime. Our results thus seem to suggest that
all third bodies in systems comparable to ours invariably tend towards
the shear-band-like state and, ultimately, phenomena like cold welding
or real shear banding if the third body density is high (ours still is
very porous). If the particles are sufficiently far apart and the
normal load is not too high, though, it seems that the rolling regime
can be maintained over long sliding distances \cite{Zanoria1993,
  zanoria1995effects, zanoria1995formation}.  One has also to consider
that the friction force will be very high in the shear-band-like
regime. This could lead to local heating and melting \cite{Chen2017}
if heat transport is not efficient enough. If melting can be avoided,
the friction force can be reduced by the formation of new, rolling
wear particles. This occurs as soon as the local contact size grows to
a critical value \cite{Aghababaei2016, Brink2021} if the third body
has a nonzero shear resistance \cite{Brink2019}. In the present work,
we can assume that the shear strength of the shear-band-like third
body has been reduced to around \SI{4}{GPa} (as in the bulk amorphous
state of the material \cite{Brink2019}) or even \SI{3}{GPa} as
suggested by Eq.~\ref{eq:sb-friction}. Then the minimum diameter for
newly formed loose wear particles \cite{Aghababaei2016, Brink2019}
would be approximately $\SI{70}{nm}$ or \SI{130}{nm}, respectively,
and therefore larger than the simulation cell. We consequently do not
observe this formation of new particles.

\section{Conclusion}

We investigated the early stages of the evolution of the third body on
the nanoscale, starting with wear particles that increasingly become
elongated cylindrical rollers and finally merge to form a porous,
shear-band-like layer. We found that the presence of wear particles
and surface roughness significantly influences friction and wear at
the nanoscale, and simple macroscopic laws are not applicable without
modification. This is in contrast to typical sphere-on-flat geometries
used in previous work \cite{Mo2009}. In the first phase (rolling
regime), Amontons--Coulomb friction is enhanced by a term proportional
to the surface roughness. A fraction of the frictional work
proportional to the surface roughness is used to grow the wear
volume. In the second phase (shear-band-like regime), pure
Bowden--Tabor-like friction is recovered, which is proportional to the
real contact area \cite{Bowden1939}. This regime exhibits a very low
effective wear rate because matter exchange takes place
bidirectionally between first and third bodies. Ultimately, third
bodies originate from the elementary, nanoscale mechanisms described
here. This understanding can provide a pathway to develop
physics-based wear models.

\section*{Acknowledgment}
Financial support by the Swiss National Science Foundation (grant
\#197152, ``Wear across scales'') is gratefully acknowledged.
Computing time was provided by a grant from the Swiss National
Supercomputing Centre (CSCS) under project IDs s784 (``The evolution
of rough surfaces in the adhesive wear regime'') and s972 (``Surface
and subsurface evolution of metals in three-body wear conditions''),
as well as by {\'E}cole polytechnique f\'ed\'erale de Lausanne (EPFL)
through the use of the facilities of its Scientific IT and Application
Support Center.

\section*{Author contributions}

T.\,B.\ and J.-F.\,M.\ designed the study, T.\,B.\ ran the simulations
and visualized the results, and T.\,B.\ and E.\,M.\ performed
analyses. All authors participated in the preparation of the
manuscript.

\appendix

\section{Choice of material}
\label{sec:mater-choice}

To simulate the formation of third bodies we needed a material that
can sustain a debris layer without welding immediately, while being
computationally affordable with regard to time and length scales.  We
started our search for suitable model materials with a range of
realistic metal potentials for Cu \cite{Mishin2001} and Al
\cite{Mishin1999}, as well as a potential \cite{Zhou2004} suitable for
Cu--Ni--Co--Fe high-entropy alloys \cite{Koch2017}. Here, we chose two
single-crystalline first bodies of size \SI{70 x 70 x 58}{nm} (around
35,000,000 atoms per simulation) with rough surfaces. The abrasive
particles quickly scratched the surface, digging into the
material. This either lead to a clumping of the particles and
suppression of rolling or a quick reduction of the gap between the
first bodies and welding. No sustained rolling regime could be
observed and friction quickly exceeded 1.0
(Fig.~\supplref{fig:welding-metal-sims}). The suppression of rolling
is likely due to a combination of the relatively low hardness of
metals with the high adhesion. We therefore also tried to use a Ni--H
potential \cite{Angelo1995}, where we manually introduced large
amounts of hydrogen into the gap of the nickel--nickel contact at
intervals of \SI{1}{ns} in order to passivate the surfaces. This did
not suppress the welding or enable continued rolling of the
particles. Since metallic glasses are known for higher hardness than
crystalline metals \cite{Schuh2007}, we also performed a test run with
a Cu$_{64}$Zr$_{36}$ glass model \cite{Cheng2009a} with first body
size of \SI{60 x 60 x 29}{nm} and initially rough surfaces. This
system quickly exhibited local surface melting despite the strong
thermostats and welded [Fig.~\supplref[(g)]{fig:welding-metal-sims}].

For the main simulations, we therefore chose to use a model potential
for a hard, silicon-like material \cite{Stillinger1985,
  Holland1998erratum}, since the above behavior has often been
observed for softer systems and metals \cite{Sorensen1996,
  Spijker2011, Pastewka2011, Stoyanov2013, Romero2014,
  Aghababaei2016}. This potential is a modification of the original
Stillinger--Weber potential \cite{Stillinger1985} which reproduces the
brittleness of the material \cite{Holland1998, Holland1998erratum,
  Hauch1999, Holland1999}. While other silicon properties are not
reproduced correctly \cite{Holland1998erratum}, the high hardness and
brittleness are optimal to observe wear at the nanoscale at an
acceptable computational cost \cite{Brink2019}. First bodies of size
\SI{58 x 58 x 23}{nm} were created (around 8,000,000 atoms in the
simulation cell). Simulations with a single-crystalline model Si in
diamond structure showed that sustained rolling of the wear particle
could be achieved. Nevertheless, the wear rate was somewhat low
(Fig.~\supplref{fig:friction-wear-big-particles}) and we wanted to
avoid the anisotropy of the single crystal. We thus chose a
nanocrystalline material created by Voronoi tessellation
\cite{Derlet2003} with average grain size of \SI{2.5}{nm}. The weak
planes introduced by the grain boundaries facilitate a higher wear
rate (Fig.~\supplref{fig:friction-wear-big-particles}) and thus less
computational expense to observe the formation of the third body. We
therefore used this material for all further investigation.

\section{Choice of initial third body, surface morphology, and
  simulation size}
\label{sec:setup-choices}

Our initial investigations of the different materials always used four
large wear particles, with diameter \SI{20}{nm} for the metals and
\SI{16}{nm} for the metallic glass and model silicon. For the
nanocrystalline system, we compared a simulation with initially flat
surface and round particles to a simulation with initially rough
surface and particles in the shape of rhombicuboctahedra. Rough
synthetic surface meshes were generated with self-affine roughness
\cite{Wu2000} using the software \textsc{tamaas} \cite{Frerot2020}. As
roughness parameters we used a Hurst exponent of 0.8, a lower
wavelength cutoff of \SI{0.5}{nm}, an upper wavelength cutoff of
\SI{11}{nm}, and an rms of heights of \SI{2}{nm}. The
continuum surface was then used to carve out the atomistic
surface. Figure~\supplref[(b)--(d)]{fig:friction-wear-big-particles}
shows that the initial friction force and wear rate of the rough
surface were higher in accordance with Eqs.~\ref{eq:rolling-friction}
and \ref{eq:wear}. After running-in, both simulations give equivalent
results.

We did not observe particle--particle interactions, though, and
decided to reduce the particle size to \SI{8}{nm} to be able to
accommodate 16 particles in the simulation cell. We used mostly
rhombicuboctahedra, but verified that the results are comparable with
initially round particles (Fig.~\supplref{fig:wear-particle-shape}).
For these simulations, we started with the rough surfaces produced by
wear with the bigger particles.

In order to speed up the simulations, we also reduced the thickness of
the bulk material in $z$ direction from \SI{23}{nm} to \SI{11}{nm} for
each one of the first bodies. We tested if the thickness influences
wear rate or friction force by repeating a sliding simulation with a
bulk thickness of \SI{6}{nm} over a sliding distance of
\SI{1}{\micro\meter}. As shown in Fig.~\supplref{fig:smaller-bulk}, we
could not detect any difference apart from the expected thermal
fluctuations. The atomic shear strain \cite{Shimizu2007} reveals that
this is because plastic events occur mostly close to the surface.

Finally, we investigated the initial placement of the wear
particles. The rolling regime could be reproduced over a comparatively
long sliding distance by having equal spacing between the wear
particles, as shown in Fig.~\ref{fig:snapshots-3rd-body-evolution}(b).
This is because adhesive bridges between the particles mostly formed
perpendicular to the sliding direction, allowing cylindrical rollers
to form. As a comparison, we also placed the particles randomly at the
start of the simulation. Due to their close spacing, adhesive bridges
in sliding direction occurred and immediately destabilized the rolling
regime (Fig.~\supplref{fig:random-particle-placement}). In reality, a
random placement of the particles would always be expected, which
means that the rolling cylinders can only remain stable if the
interparticle distance remains large during sliding.

\section{Verification simulations}
\label{sec:verif-sims}

While the sliding velocity is on the order of experimentally
achievable velocities, it is nevertheless quite high. We investigated
if there are any obvious rate effects by taking a snapshot of the
sliding simulation with $F_N = \SI{7.69}{\micro\newton}$ after
$s = \SI{1}{\micro\meter}$ and slowing the sliding velocity down to
zero over a time of \SI{5}{ns}. After this, we kept the system static
for \SI{5}{ns} and afterwards ramped up the velocity again over
\SI{5}{ns}. Figure~\supplref{fig:ramp-down-up} shows that the friction
force as a function of sliding distance is not strongly affected by
the velocity in this setup. This indicates that the friction force is
more sensitive to the surface morphology than to dynamic effects.

The use of rigid particles resembles the abrasive wear case of
initially introducing much stiffer and harder wear particles into the
system. Here, we end up with a behavior that is closer to abrasive
wear due to strong adhesion and the coating of the particles with bulk
material. We therefore did short test simulations (for a sliding
distance of \SI{10}{nm}) in which we made the initial wear particles
non-rigid and let them deform fully according to the interatomic
forces. One of the simulations was restarted from the original run
after $s = \SI{1}{\micro\meter}$ (in the rolling regime) and the other
after $s = \SI{3}{\micro\meter}$ (in the shear-band-like regime).
Figure~\supplref{fig:non-rigid} shows that this does not lead to
deviations of the friction force. In the rolling regime, no plasticity
was observed inside the originally rigid particles, while the
shear-band-like regime exhibited plasticity all over the third
body. We therefore expect that the latter will lead to welding more
quickly without the rigid particles.

\balancecolsandclearpage


\section*{Supplementary material}

# Effect of wear particles and roughness on nanoscale friction

Tobias Brink, Enrico Milanese, and Jean-François Molinari

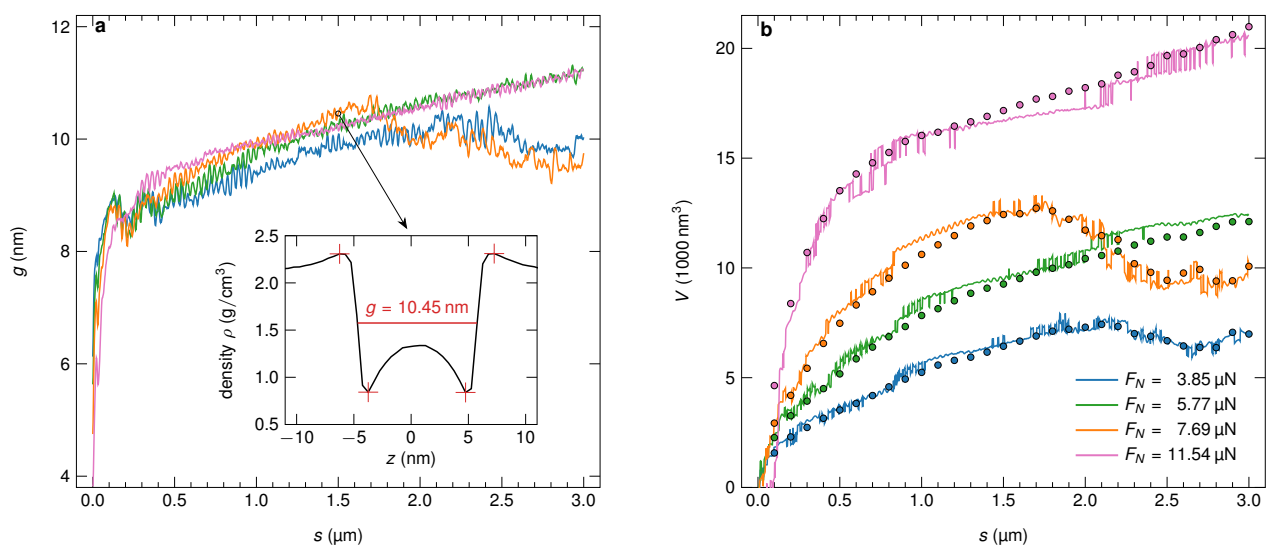

**Figure S1:** Tribolayer thickness and wear volume calculated from density profiles. **a**, Thickness $g$ of the tribolayer. The continued growth in the rolling regime is in accordance with the agglomeration of mass in the rolling wear particles. It can be seen that the tribolayer width stagnates together with the wear rate in the shear-band-like regime after around $s = 2\,\mu m$ for $F_N = 3.85\,\mu N$ and $7.69\,\mu N$. Inset: Calculation of $g$ from a density profile in $z$ direction (normal to the surfaces). **b**, Wear volume calculated by integrating over the density profile (lines) compared to the computation using the number of atoms in the tribolayer after separation (data points). The lines are shifted to zero at the beginning of the sliding simulation since the integrals include the initial wear particles, which we do not count in the wear volume.

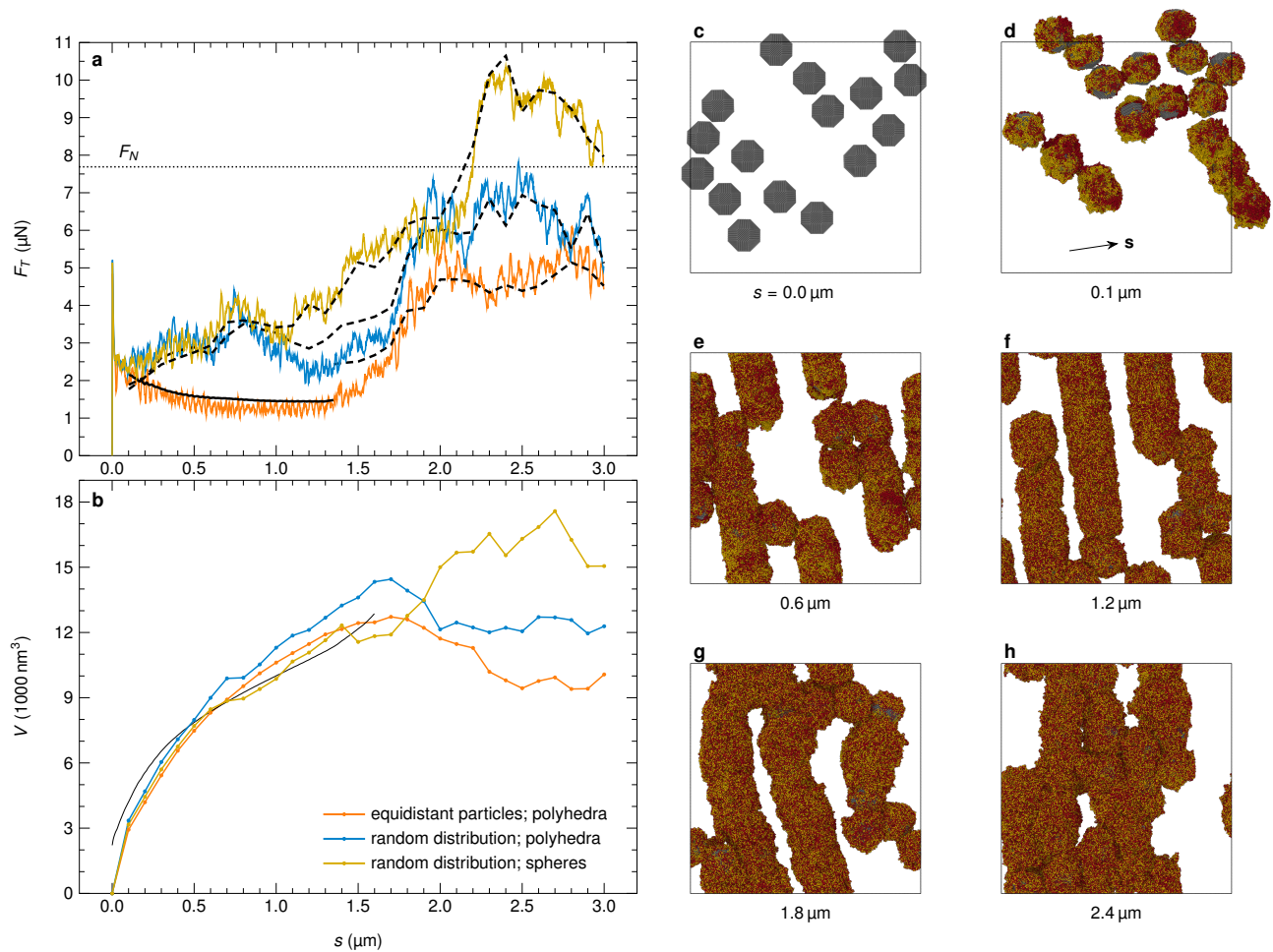

**Figure S2:** Influence of the initial particle placement. **a**, Compared to the regular arrangement with equidistant particles used for the main results (orange), random initial particle placement leads to an immediate transition to a high-friction regime. Solid black lines correspond to Eq. 1 in the main text, dashed black lines to Eq. 3. **b**, The wear rate does not immediately stagnate, though, as might be expected for the shear-band-like regime. A look at the morphology of the third body (panels **c–h**) can explain this: Parts of the system still clearly consist of rolling cylinders that continue to grow and pick up material. The wear rate only stagnates when all cylindrical particles strongly interact and/or merge (panels **g** and **h**). The solid black line in panel **b** corresponds to Eq. 6 in the main text, computed for the equidistant particles in the rolling regime.

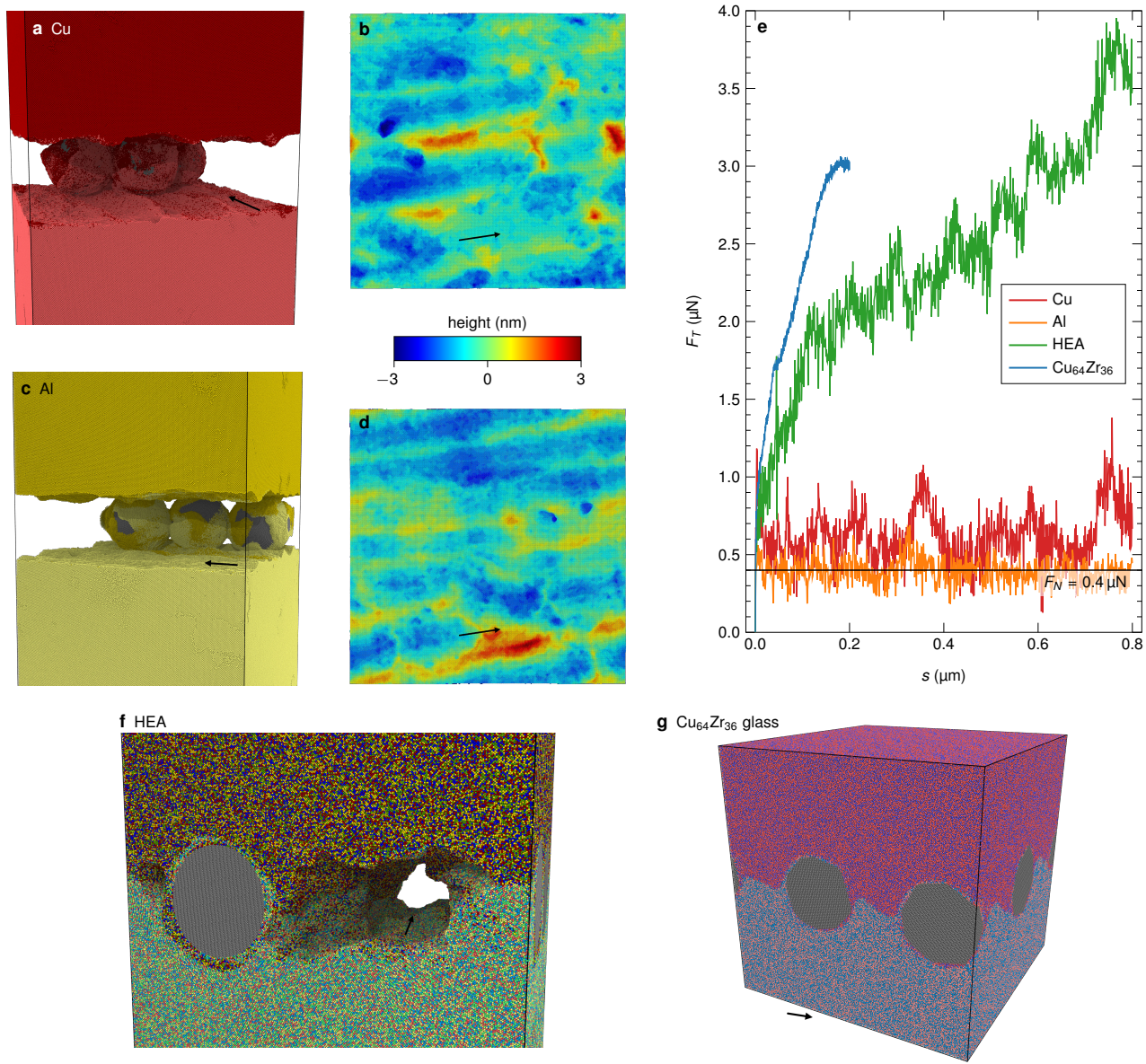

**Figure S3:** Sliding simulations on different metals. Copper (panels **a** and **b**) and aluminum (panels **c** and **d**) do not exhibit sustained rolling of the wear particles, but the particles stick together almost immediately, leading to scratching (see height maps in panels **b** and **d**) and friction coefficients of one or more (panel **e**). **f**, The high-entropy alloy shows more adhesion and plastic flow than the elemental metals, leading to large pickup of material and very quick welding. **g**, The metallic glass welds instantly due to strong softening of the surface.

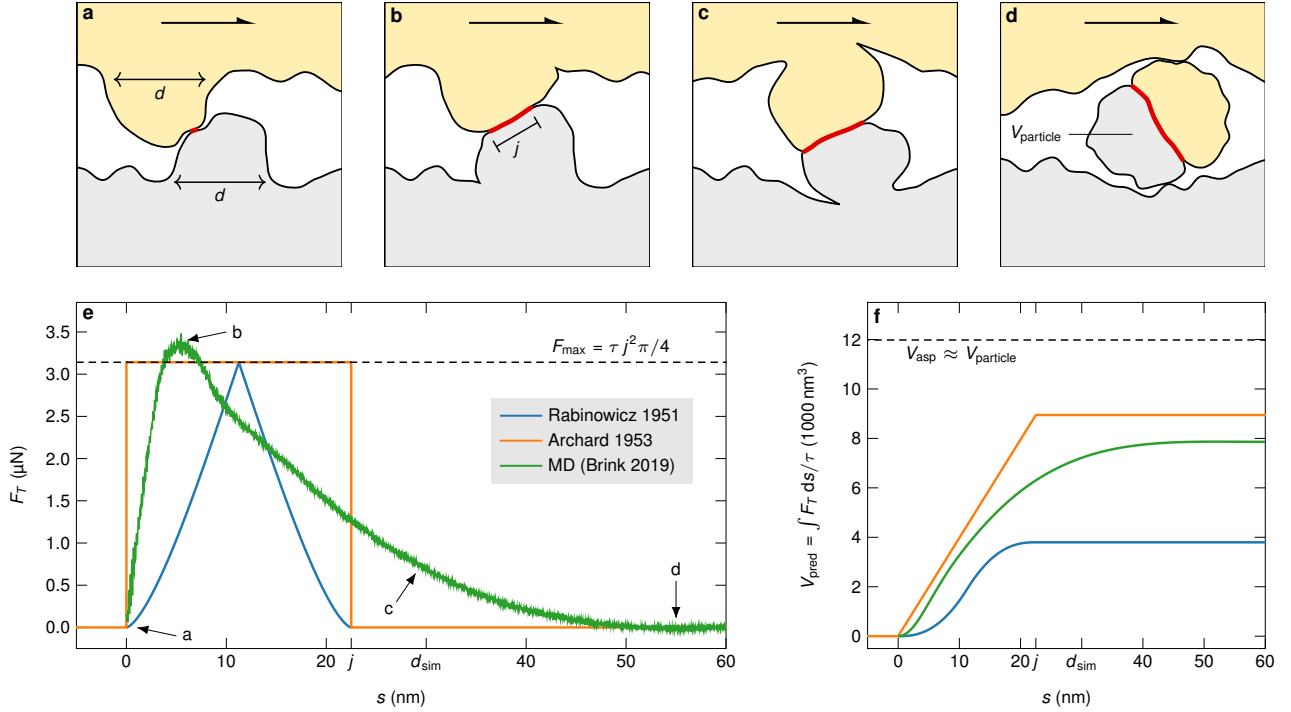

**Figure S4:** A single-asperity wear law. In the case of wear particle formation with significant plastic activity, it has been found that the wear volume in the single-asperity case is related to the tangential work by [1] $V = \int_0^s F_T \mathrm{d}s / \omega\tau \approx F_T s / \omega\tau$ (Eq. 5 in the main text), with $\omega$ being a shape factor close to unity. This single-asperity wear law can be derived by assuming that the asperity is loaded to its shear limit $\tau$ upon contact (panel **a**). During sliding, the particle starts to detach (panel **b** and **c**) and the full detachment (panel **d**) will take place at a sliding distance of $s = d$, where $d$ is the asperity diameter, at which point $F_T$ becomes negligible, allowing the approximation of a constant tangential force. The shear limit can be converted to the tangential force via $F_T = \omega\tau d^2$, assuming that the contact area is given by the asperity diameter ($j \approx d$). Measuring only the work $W = \int_0^s F_T \mathrm{d}s \approx F_T s$, one can estimate the asperity volume (and therefore the wear particle volume) via $s = d$ and $W/\tau = F_T d / (F_T / \omega d^2) = \omega d^3$. In this simple picture, all dissipation goes into forming a wear particle. In multi-asperity systems, additional frictional dissipation mechanisms occur, wherefore a wear coefficient $k$ is introduced that takes into account that not all the tangential force is used for wear. **e**, Taking data from an earlier simulation study [2], we can compare the simple models of the tangential force profile by Rabinowicz [3] and Archard [4] with a particle-detachment simulation of a brittle material. As already indicated in panel **b**, the contact area is given by a contact diameter $j < d$, which we use here instead of $d$. **f**, We can also see that the measured wear volume (dashed line) exceeds the predictions, meaning that the wear process is more efficient. The reason is that a brittle asperity is not loaded to its shear limit in all of its volume, since brittle fracture can obviously occur before reaching the elastic limit. Nevertheless, the approximation reproduces the correct order of magnitude [1], as found before.

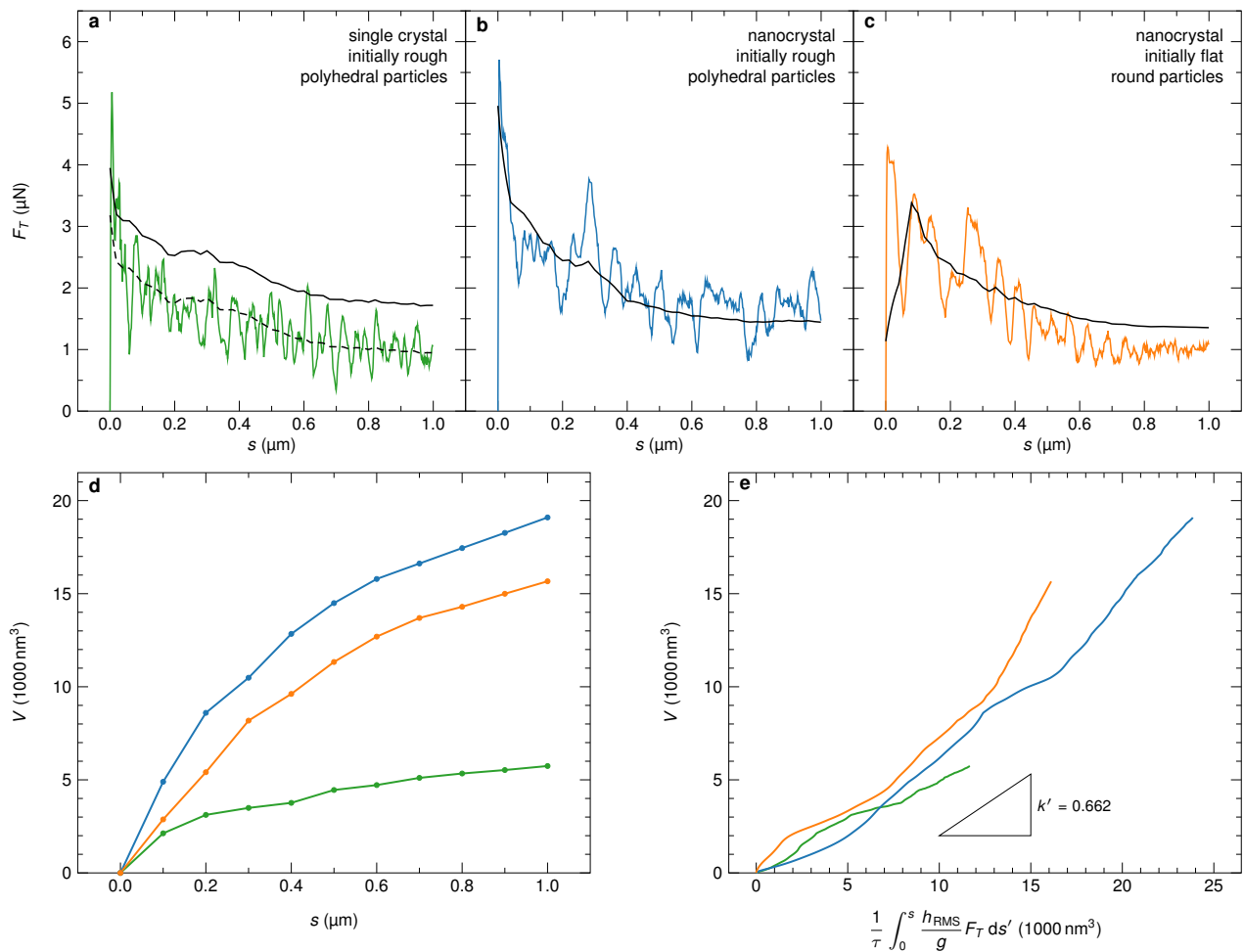

**Figure S5:** Friction and wear data of the same material used in the main text with four, large initial wear particles. The normal force was $F_N = 7.69\,\mu\text{N}$. **a,** For the single-crystalline first bodies, the friction force (green line) is reduced compared to the prediction from the model in Eq. 1 in the main text (solid, black line). A fit (dashed, black line) can be achieved by reducing $\mu_0^r$ to 0.03. This indicates that the non-roughness contribution to the friction is lower, possibly due to a reduction of the contact area of the rollers (higher hardness of the single crystal) or due to a reduced effective shear strength (cf. Eq. 2 in the main text). The nanocrystalline first bodies starting from initially rough (panel **b**) and initially flat (panel **c**) surfaces follow the prediction using Eq. 1 in the main text (solid, black lines). **d,** In terms of wear volume, the single-crystalline system wears more slowly and the wear rates of both nanocrystalline systems are similar at the end. As expected, the initial wear rate for the rough surface is higher than for the flat surface. **e,** When plotting the wear volume over the wear prediction from Eq. 6 in the main text (cf. Fig. 4d in the main text), the curves collapse and we obtain approximately the same wear coefficient $k'$ as before.

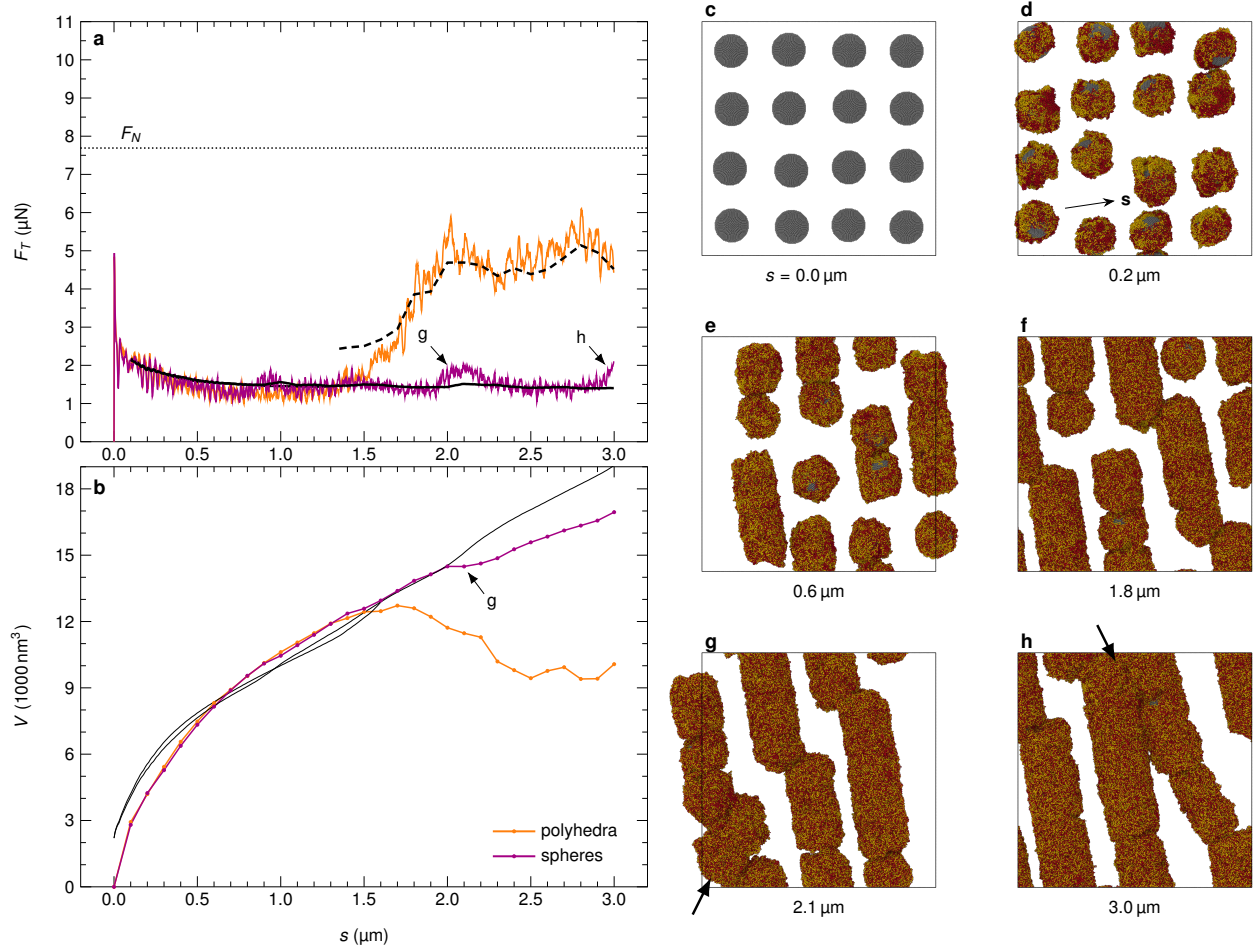

**Figure S6:** Influence of initial wear particle shape. We tested both polyhedral and round particles, but found that this does not significantly change either the friction (panel **a**) or the wear rate (panel **b**). This is because the particles are quickly coated by material picked up from the first bodies, resulting in the same shape after a short time in both cases. Solid black lines in panel **a** correspond to Eq. 1 in the main text, dashed lines to Eq. 3, and solid black lines in panel **b** correspond to Eq. 6. The only difference between the simulations is when the transition to the shear-band-like regime with high friction and low wear rate occurs. This transition is a stochastic process, though, and thus not a result of the initial particle shape. Panels **c**–**h** show the evolution of the third body starting from round particles. The rolling regime is retained throughout the simulation. Spikes in the friction can be connected directly to events where the rolling cylinders interact (panels **g** and **h**). Indeed, the transition to the shear-band-like regime seems to initiate already partially at around $s = 2\,\mu m$, where the wear rate becomes lower than the prediction for the rolling regime (panel **b**) and the particles start sticking together (panels **g** and **h**).

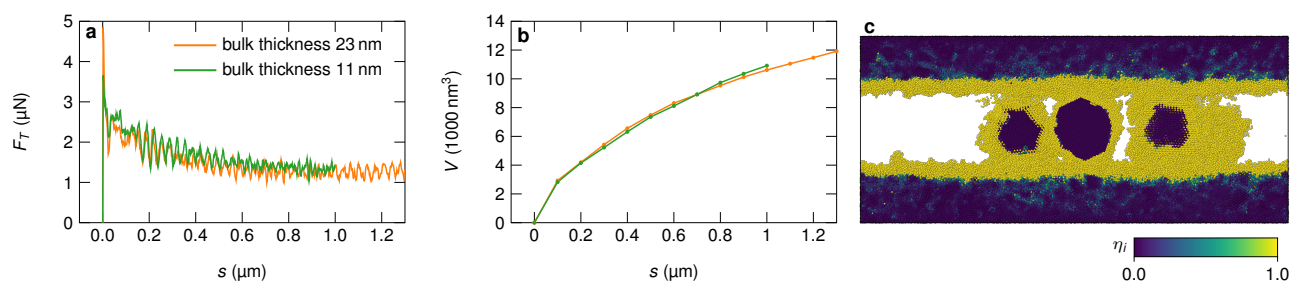

**Figure S7:** Verification that reducing the initial bulk height does not influence friction or wear within the simulated timescale. Since the system sizes in MD simulations are constrained due to the high computational demands of the method, finite size effects must be excluded. Here, we show that the friction (panel **a**) and wear (panel **b**) response of a system that is much smaller than those in the main results is the same. **c**, This can be explained by the fact that the plasticity, here visualized as the von Mises invariant $\eta_i$ of the total atomic shear strain [5] at $s = 3\,\mu m$ for the sample with bigger initial bulk thickness, is quite localized at the surface.

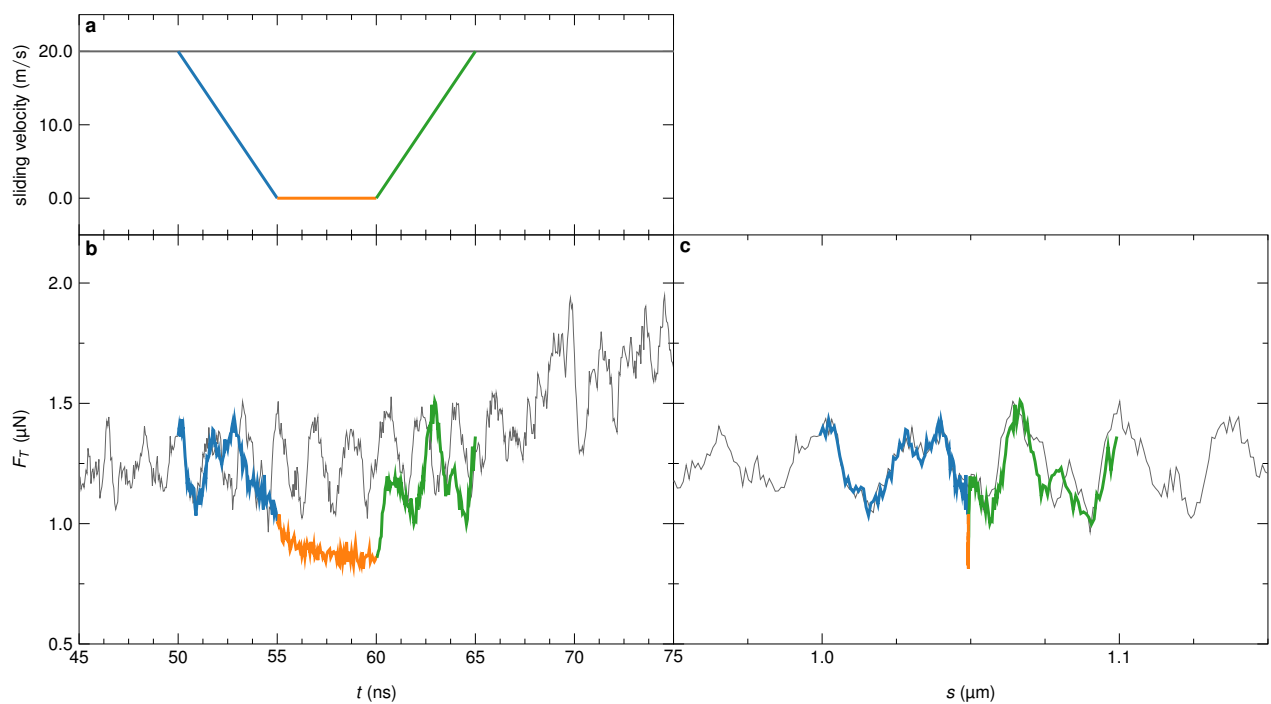

**Figure S8:** Change of the sliding velocity. **a**, We investigated the influence of the sliding velocity by restarting the sliding simulation with $F_N = 7.69\,\mu N$ at $s = 1\,\mu m$ but then slowly ramping down the velocity from $20\,m/s$ to $0\,m/s$ (blue), holding for $5\,ns$ (orange), and ramping up the velocity again afterwards (green). Grey curves indicate the original, constant velocity simulation. **b**, The friction force relaxes during the holding period, but picks up again quickly afterwards. **c**, When plotting the same data over the sliding distance, it becomes clear that the friction response is not sensitive to the sliding velocity, but to the surface morphology.

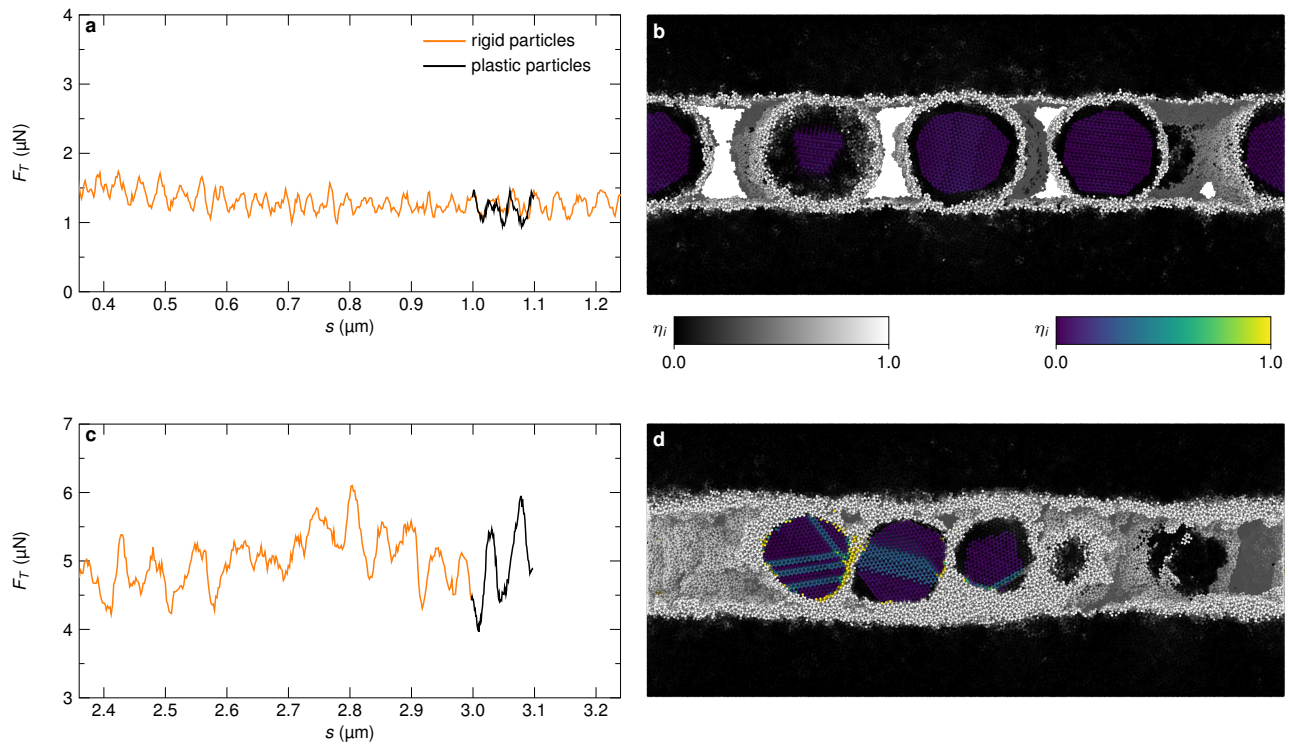

**Figure S9:** Making the abrasive particles non-rigid. **a**, Restarting the simulation with $F_N = 7.69\,\mu\text{N}$ at $s = 1\,\mu\text{m}$ in the rolling regime with non-rigid particles does not lead to any change in the friction force response. **b**, The atomic shear strain [5] (blue to yellow: originally rigid particles, black to white: rest of the system) shows no plasticity in the core of the wear particles. **c**, When continuing the simulation at $s = 3\,\mu\text{m}$, the friction force is also not affected by the rigidness of the particles. **d**, Nevertheless, plasticity can be observed inside the particles in accordance with the fact that the tribolayer is in a shear-band-like state. Here, shear forces get transmitted through the whole third body.

## Supplemental references



\end{document}